\documentclass{svproc}
\usepackage{graphics}
\usepackage{makecell, cellspace, caption}
\usepackage{color, colortbl}
\usepackage{xcolor}
\usepackage{multirow}

\usepackage{graphicx}
\usepackage{tabularx,booktabs}
\usepackage{dingbat}
\usepackage{diagbox}

\usepackage{epsf,graphicx}
\usepackage{latexsym,amssymb}
\usepackage{cite}
\usepackage{subfig}
\usepackage{url}
\usepackage{amsmath}
\usepackage{rotating}
\usepackage{amsmath}
\usepackage{subfig}
\usepackage{array}

\usepackage{multirow}
\usepackage{textcomp}

 \usepackage{lscape}
 \usepackage{longtable}
\usepackage{multirow}
\usepackage{float}
\usepackage{adjustbox}
\usepackage{amsmath,amssymb,amsfonts}
\usepackage{graphicx}
\usepackage{booktabs}
 \usepackage{url}
\usepackage{makecell}
\usepackage{float}
\usepackage{adjustbox}
\usepackage{lineno,hyperref}
\modulolinenumbers[5]
\usepackage{mathptmx}
\usepackage{longtable}
\usepackage{multirow}
\usepackage{graphicx}
\usepackage{diagbox}
\usepackage{dsfont}
\usepackage{rotating}
\usepackage[section]{placeins}
\usepackage{graphicx}
\usepackage{relsize}

\usepackage{pdflscape}
\usepackage{cases}
\usepackage{afterpage}
\usepackage{textcomp}
\usepackage{array,multirow,graphicx}
\usepackage{capt-of}

\begin{document}
\mainmatter  
\title{Voting Classifier-based Intrusion Detection for IoT Networks}
\titlerunning{}  
%
\author{Muhammad Almas Khan\inst{1} \and Muazzam A Khan\inst{1} \and Shahid Latif\inst{2}\and Awais Aziz Shah\inst{3}\and Mujeeb Ur Rehman\inst{4}\and Wadii Boulila\inst{5,6} \and Maha Driss\inst{5,6} \and  Jawad Ahmad\inst{7} }
\authorrunning{Muhammad Almas Khan et al.} 
%
\tocauthor{Ivar Ekeland, Roger Temam, Jeffrey Dean, David Grove,
	Craig Chambers, Kim B. Bruce, and Elisa Bertino}
\institute{Quaid-i-Azam University, Islamabad, Pakistan\and
	Fudan University, Shanghai, China\and
	Polytechnic University of Bari, Italy \and
	Riphah International University, Islamabad, Pakistan\and
    University of Manouba, Manouba, Tunisia\and
	 Taibah University, Medina, Saudi Arabia \and 
	 Edinburgh Napier University, Edinburgh, UK
	}

\maketitle              

\begin{abstract}
Internet of Things (IoT) is transforming human lives by paving the way for the management of physical devices on the edge. These interconnected IoT objects share data for remote accessibility and can be vulnerable to open attacks and illegal access. Intrusion detection methods are commonly used for the detection of such kinds of attacks but with these methods, the performance/accuracy is not optimal. This work introduces a novel intrusion detection approach based on an ensemble-based voting classifier that combines multiple traditional classifiers as a base learner and gives the vote to the predictions of the traditional classifier in order to get the final prediction. To test the effectiveness of the proposed approach, experiments are performed on a set of seven different IoT devices and tested for binary attack classification and multi-class attack classification. The results illustrate prominent accuracies on Global Positioning System (GPS) sensors and weather sensors to 96\% and 97\% and for other machine learning algorithms to 85\% and 87\%, respectively. Furthermore, comparison with other traditional machine learning methods validates the superiority of the proposed algorithm.
\end{abstract}
\section{Introduction}
Over the past few years, the Internet of Things (IoT) has witnessed rapid growth and adaptability in every sector of life \cite{b1}, \cite{b2}, \cite{b3}, \cite{hajjaji2021big}, and \cite{b5}. Figure 1 shows the domain of IoT/IIoT in real life. With the rapid growth of IoT, the security of the smart systems based on IoT is an important task. Current literature reveals the importance of cyber-attack detection schemes for smart systems \cite{churcher2021experimental, shafique2021detecting, ali2020network, qayyum2020chaos,masood2020novel}. History shows that an IoT device compromised by an attacker resulted in a power blackout and 225000 people were affected due to the lower quality of security mechanism \cite{b7}. IoT technology has many features such as inter-dependency among devices, constrained, diversity, etc \cite{b10}. Understanding theses features in detail help us in securing the smart systems from upcoming threats. Out of these features identified in \cite{b10}, for example, the interdependence of IoT devices provide less human involvement and takes a smart decision when one device implicitly controls another within the network as they share data with each other. To understand the inter-dependence, let us considers a simple scenario of smart home devices, where the thermostat device senses the home temperature and compares it with the threshold value. If the temperature is out of the threshold stored in the thermostat, it tries to balance the environmental temperature. In order to balance the environmental temperature it checks the Air Condition (AC) smart plug. If the AC is unplugged, it automatically opens the windows in order to stabilize the environmental temperature and allow ventilation. Without proper security, this interdependence among IoT devices provides an opportunity to intruders by compromising an IoT device they can open the door or window of a facility. For instance, the code injection and Man in the Middle (MitM) attack \cite{zolanvari2019machine} can compromise an IoT device.\begin{figure}[ht!] 
\centering
\includegraphics[width=3in]{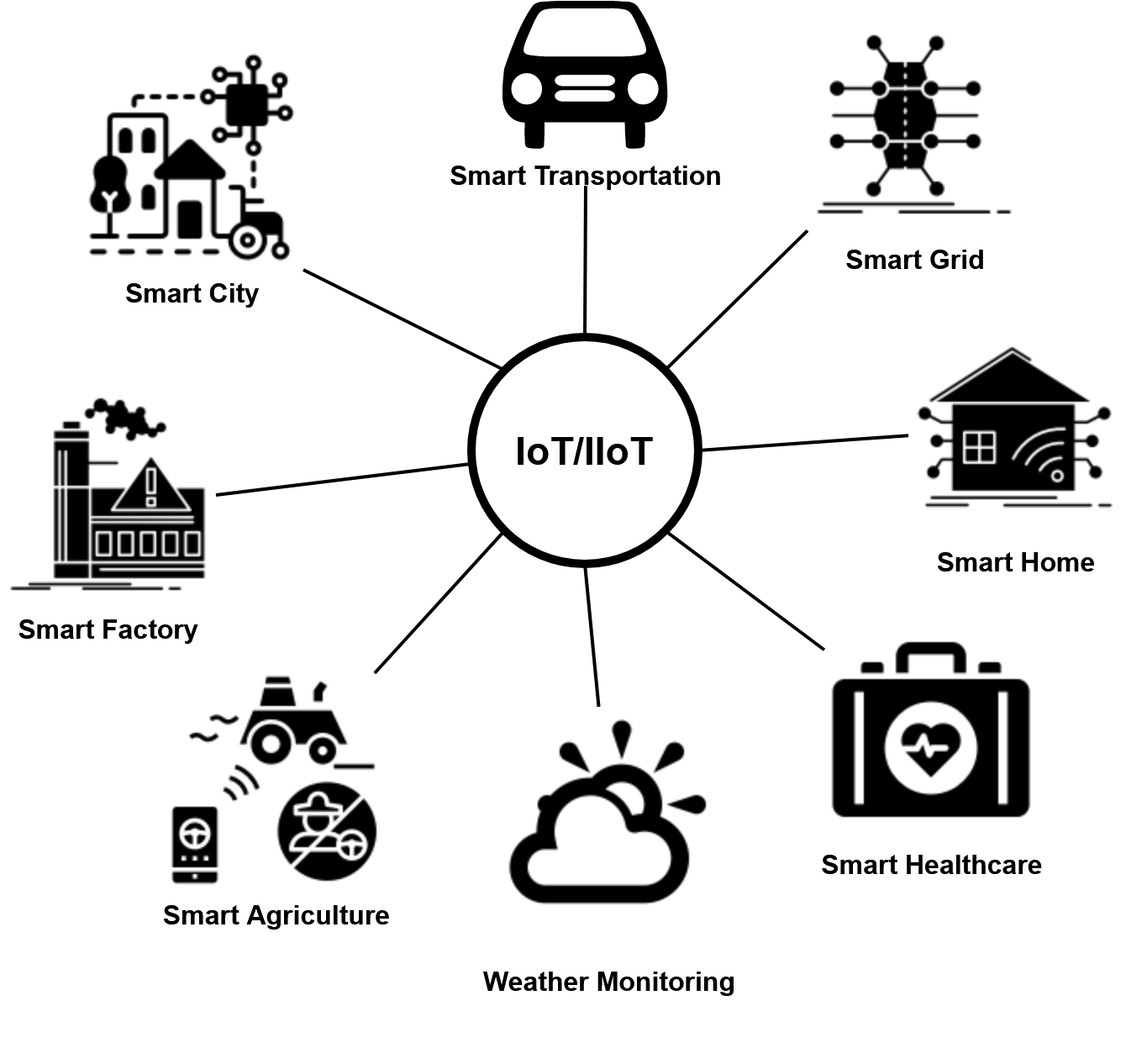}
\caption{Internet of Things Applications}
\label{Courant_2}
\end{figure} In code injection, the adversary injects malicious code to IoT devices and performs alteration of the data. Similarly, in MitM attack the intruder eavesdrops and tries to communicate messages between two nodes. Here, the attacker can control the action of other node. To secure each IoT device within the IoT network and make secure the interdependence among devices, this paper evaluates the ensemble-based voting classifier for intrusion detection in the IoT network. Ensemble-based voting classifier combined the traditional ML algorithm and give voting to its prediction to get the final prediction by the ensemble-based voting classifier. There are two types of voting e.g soft voting and hard voting. A detailed mathematical description of the proposed algorithm is given in section III. The performance of the conceived algorithm is evaluated on the latest dataset which is a real representative of IoT networks called Ton-IoT \cite{b12}. Ton-IoT \cite{b23} has been selected as the dataset in our experiments after analyzing the available datasets in the current state of the art representing a real IoT/IIoT network. The novel contributions of this paper are: 
\begin{itemize}
    \item Proposal of an ensemble-based model for attack classification.
    \item Evaluation of the proposed model for binary and multi-attacks class IoT dataset. 
     \item Comparison of the proposed model with traditional machine learning algorithms using evaluation metrics such as  accuracy, precision, recall, and F1-score.
115 F1-measure. 
   \end{itemize}
The remainder of the paper has been organized as follows: Section II presents background and detailed discussion on the related work and existing threats in IoT. The problem statement and proposed solution is discussed in section III. Section IV describes the selection and pre-processing of the selected datasets. Section V presents the cross-comparison between the existing state of the art intrusion detection algorithms and our proposed methodology. Finally, Section VI concludes the work and draws some potential future directions.

\section{Literature Review}
IoT is an innovative technology for making the systems smart by automating the working environment to reduce human involvement and improve system performance \cite{batool2019identification}. Security in IoT is one of the main concerns because innovative types of cyberattacks are emerging every day as IoT technology is rapidly evolving \cite{b10}. Without security, these networks are vulnerable to attacks and a favorite place for intruders. To attack and take control over the whole IoT system, attackers usually compromise an IoT device and control indirectly the other devices \cite{b7} \& \cite{b8}. To prevent these attacks at early stages, one of the possible ways is to detect the intrusion or malicious activity of an attacker on the network. In this regard, Intrusion detection algorithms play an important role in detecting such malicious activities within the networks. Table I provides the summary of this section. In \cite{b13}, 4 ensemble techniques such as  Boosted Trees, Bagged Trees, Subspace Discriminant, and RUSBoosted Trees were implemented to propose RPL based network intrusion detection for IoT networks. The proposed NIDS are evaluated on RPL-NIDDS17 dataset. To prevent the malicious events in IoT networks, more specifically the botnet attack against MQTT, HTTP, and DNS protocols, \cite{b14} proposed an ensemble-based intrusion detection technique. Three machine learning algorithms namely DT, NB, and ANN were combined to develop an adaptive boost ensemble method for attack detection. The introduced scheme was evaluated on two datasets namely UNSW-NB15 and NIMS botnet dataset.\begin{table}[!ht]
\centering
\caption{Intrusion detection models with evaluation on datasets and evaluation metrics}
\label{tab:my-table}
\resizebox{\textwidth}{!}{%
\begin{tabular}{clccc}
\hline
\multicolumn{2}{|c|}{\textbf{Author \& paper title}} & \multicolumn{1}{c|}{\textbf{Model}} & \multicolumn{1}{c|}{\textbf{Evaluation Dataset}} & \multicolumn{1}{l|}{\textbf{Evaluation metrics}} \\ \hline
\multicolumn{2}{|c|}{\textbf{\cite{b13}}} & \multicolumn{1}{c|}{Ensemble learning} & \multicolumn{1}{c|}{\textit{RPL-NIDDS17}} & \multicolumn{1}{c|}{Accuracy, ROC} \\ \hline
\multicolumn{2}{|c|}{\textbf{\cite{b14}}} & \multicolumn{1}{c|}{Adaptive boost} & \multicolumn{1}{c|}{\textit{\begin{tabular}[c]{@{}c@{}}UNSW-NB15\\ NIMS\end{tabular}}} & \multicolumn{1}{c|}{\begin{tabular}[c]{@{}c@{}}Accuracy,  False positive rate\\ Detection rate, ROC\end{tabular}} \\ \hline
\multicolumn{2}{|c|}{\textbf{\cite{b15}}} & \multicolumn{1}{c|}{Ensemble Voting} & \multicolumn{1}{c|}{\textit{\begin{tabular}[c]{@{}c@{}}CIC-IDS2017\\ NSL-KDD\\ AWID\end{tabular}}} & \multicolumn{1}{c|}{\begin{tabular}[c]{@{}c@{}}Accuracy, Precision\\ Detection rate, F-measure\end{tabular}} \\ \hline
\multicolumn{2}{|c|}{\textbf{\cite{b16}}} & \multicolumn{1}{c|}{XGBoost} & \multicolumn{1}{c|}{\textit{NSL-KDD}} & \multicolumn{1}{c|}{\begin{tabular}[c]{@{}c@{}}Accuracy, Precision\\ Recall, F-measure, ROC\end{tabular}} \\ \hline
\multicolumn{2}{|c|}{\textbf{\cite{b17}}} & \multicolumn{1}{c|}{XGBoost} & \multicolumn{1}{c|}{\textit{UNSW-NB15}} & \multicolumn{1}{c|}{Accuracy} \\ \hline
\multicolumn{2}{|c|}{\textbf{\cite{b19}}} & \multicolumn{1}{l|}{Ensemble learning} & \multicolumn{1}{c|}{\textit{AWID}} & \multicolumn{1}{c|}{\begin{tabular}[c]{@{}c@{}}Accuracy, Precision\\ Recall,  F1-Measure\end{tabular}} \\ \hline
\multicolumn{2}{|c|}{\textbf{\cite{b23}}} & \multicolumn{1}{c|}{\begin{tabular}[c]{@{}c@{}}LR, LDA\\ kNN, RF, CART\\ NB, SVM, LSTM\end{tabular}} & \multicolumn{1}{c|}{Ton-IoT} & \multicolumn{1}{c|}{\begin{tabular}[c]{@{}c@{}}Accuracy, Precision\\ Recall, F-measure\end{tabular}} \\ \hline
\multicolumn{1}{l}{} &  & \multicolumn{1}{l}{} & \multicolumn{1}{l}{} & \multicolumn{1}{l}{}
\end{tabular}%
}
\end{table} In \cite{b15}, 2 step intrusion detection system were proposed. In the first step, the proposed method selects optimal features and in the second step, they combine machine learning algorithms namely C4.5, RF, and Random Forest by Penalizing Attribute(RF PA) to develop a voting ensemble classifier. The proposed method was evaluated on NSL-KDD, AWID, and CIC-IDS2017 datasets with evaluation metrics Accuracy, Precision, Detection rate, F-measure. In \cite{b16} Intrusion detection using XGBoost model proposed. The proposed model was evaluated on NSL-KDD dataset with evaluation metrics accuracy, precision, recall, and F1-measure. To detect intrusion on network-level \cite{b17} proposed Extreme Gradient Boosting (XGBoost) based intrusion detection, the proposed system was evaluated on the UNSW-NB15 with accuracy as an evaluation metric. For supporting the narrowband and broadband IoT applications the wifi is needed in some place, to protect the attacks on wifi, \cite{b19} proposed an ensemble-based intrusion detection by evaluated on AWID dataset with evaluation metrics accuracy, precision, recall, and F1-measure. In \cite{b31}, Liu et all studied 11 algorithms comprising 7 supervised and 3 unsupervised algorithms to find the best solution for the detection of intrusion in the IoT networks. The findings suggest XGBOOST as the best performer and Expectation-Maximization (EM) from unsupervised algorithms. The performance of 11 algorithms was tested on NSL-KDD datasets with Accuracy, Matthews correlation coefficient (MCC), and  Area Under the Curve (AUC) as evaluation metrics.

However, the evaluation performed using the datasets \cite{amin2017accelerated} plays the main role in intrusion detection. In \cite{b23}, the authors proposed a new generation telemetry dataset of IoT 4.0 which has been generated in a diverse attack scenario. This dataset contains data of 7 IoT devices having real sensor measurements. The performance of ML and DL algorithms is tested in which DT and RF outperform when their evaluation metrics such as accuracy, precision, recall, F-measure, etc were observed. However, the results show that the performance of a single ML algorithm change as the data from the sensors changes. Therefore ensemble-based learning is required in order to give optimal performance in attack detection on every sensor.
Designing an optimal intrusion detection system requires a realistic dataset as mentioned before, which is close to real-time scenarios \cite{b21, b22, b23}. In this research work, several publically available datasets are explored and evaluated based on several comparison metrics such as diverse attack scenarios, telemetry data of IoT, and availability of separate datasets for each IoT object, etc. For the designing and analyzing of an IDS, publically available datasets are KDDCUP9, NSL-KDD, Labelled Wireless Sensor Network Data Repository (LWSNDR), UNSW-NB15, Aegean Wi-Fi Intrusion Dataset (AWID), ISCX, UNW-IoT trace, UNSW-IoT, BoT-IoT, RPL-NIDDS17, CICIDS2017, and Ton-IoT. A newly generated dataset known as TON\_IoT \cite{b23} is a publicly available for the evaluation of an IDS in IoT and IIoT networks. This dataset has been generated in diverse attack scenarios and data is gathered from the Telemetry data of IoT/IIoT services. The dataset contains 7 IoT representing Fridge sensor, Garage door, GPS (Global Positioning Sensor) sensor, Modbus, Weather, Motion light sensor, and Thermostat. The data stored in these datasets are different from each other and hence obtained dataset is from heterogeneous sources. All IoT devices do not deal with the same type of data, For example, garage door IoT devices only deal with 'ON' or 'OFF' representing the state of the door. Similarly, some devices deal with real-valued numeric data. That's why the performance of the traditional ML algorithm does not remain the same with changes in the type of data. Due to the aforementioned reason, we have combined a few ML algorithms such as DT, NB, RF, and K-NN to propose an accurate classifier that could deal with any type of data with optimal performance on a maximum number of devices within IoT networks. Table 2 summarizes the publicly available dataset for analysis of an IDS. One can see from this table that the Ton-IoT dataset provides an accurate environment (based on different/separate IoT device data) for an IDS for IoT devices.
\begin{table}[]
\caption{\textbf{Comparison of publicly available datasets.} }
\label{tab:comparison}
\begin{tabularx}{\textwidth}{@{}l*{12}{c}c@{}}
\midrule
\midrule
\diagbox[width=11em]{\textbf{Comparison} \\
\textbf{Metrics}}{\textbf{Dataset}} &
\rotatebox[origin=c]{80}{KDDCUP99} & \rotatebox[origin=c]{80}{NSL-KDD} & \rotatebox[origin=c]{80}{LWSNDR}& \rotatebox[origin=c]{80}{UNSW-NB15}  & \rotatebox[origin=c]{80}{AWID}   & \rotatebox[origin=c]{80}{ISCX}  & \rotatebox[origin=c]{80}{UNW-IoT trace}   & \rotatebox[origin=c]{80}{UNSW-IoT}  & \rotatebox[origin=c]{80}{BoT-IoT}  & \rotatebox[origin=c]{80}{RPL-NIDDS17}  &
\rotatebox[origin=c]{80}{CICIDS2017}  &
\rotatebox[origin=c]{80}{Ton-IoT} \\ 
\midrule
Attack label & \checkmark & \checkmark  &
\checkmark & \checkmark & \checkmark   & \checkmark & N/A & \checkmark & \checkmark & \checkmark & \checkmark & \checkmark \\ \\
\midrule
Diverse attack scenario & x & x & x & \checkmark & \checkmark  & \checkmark & x  & \checkmark  & \checkmark & \checkmark & \checkmark & \checkmark \\ \\
\midrule
Telemetry data of IoT & x & x & \checkmark & x & x & x  & x & x & x & x & x & \checkmark\\ \\
\midrule
Separate dataset for each IoT & x & x  & x & x & x  & x & x & x & x & x & x & \checkmark \\  \\
\midrule
 Year & \rotatebox[origin=c]{50}{1998} & \rotatebox[origin=c]{40}{1998}  & \rotatebox[origin=c]{40}{2010}  & \rotatebox[origin=c]{40}{2015} & \rotatebox[origin=c]{40}{2015} & \rotatebox[origin=c]{40}{2017}& \rotatebox[origin=c]{40}{2018} & \rotatebox[origin=c]{40}{2019}  & \rotatebox[origin=c]{40}{2018}    & \rotatebox[origin=c]{40}{2018}   & \rotatebox[origin=c]{40}{2017} & \rotatebox[origin=c]{40}{2020}  \\ \\
\bottomrule
\end{tabularx}
\end{table}
\section{Proposed Ensemble-based Intrusion Detection Scheme}
The classifiers based on ensemble technique is one of the methods for generating a powerful classifier that has higher classification accuracy as compared to traditional ML classifiers. Due to hybrid in nature, ensemble-based techniques generally perform better on most of the dataset and hence give optimal results when compared to an individual traditional ML classifier. However, it is tedious to find the best pair of machine learning classifiers to the ensemble for particular data to get the best possible results \cite{b36}. In this research work, we have combined DT, NB, RF, and kNN using a voting-based technique. To understand fully the proposed method, we have outlined the mathematical description in this section.
To understand the proposed ensemble-based classifier, preliminary mathematical knowledge has been explained in this section. Prior to the mathematical details, we have highlighted an ensemble-based voting classifier in Figure 2. $C1$, $C2$ ... $C_m$ shows classifiers while $P1$, $P2$, ... $P_m$ show the predictions, respectively. To show the superiority of an ensemble-based classifier, authors have discussed different factors in \cite{boulila2009improving,al2020ensemble, b35,b36,b37}. 
\begin{figure}[ht!] 
\centering
\includegraphics[width=3in]{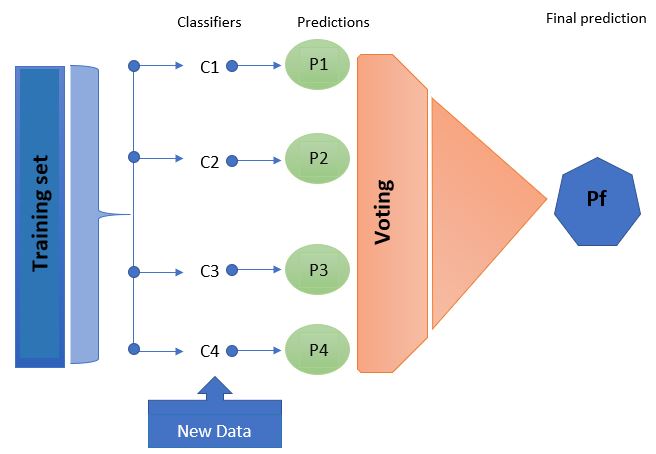}
\caption{Ensemble-based voting Classifier}
\label{Courant_2}
\end{figure}
In a voting method for binary classification, the simple majority is used while for multi-classification it is extended to plurality voting. Mathematically ensemble-based voting is written as:
\begin{equation}
  \hat{y}=arg max_{i}\left[  \sum_1^m W_{j} \chi_{A}(C_{j}(X)=i)  \right]   
\end{equation}
where the $\hat{y}$ is the prediction of ensemble model, $C_{j}$ is the classifiers ensembles, $\chi_{A}$ is the characteristic functions, $A$ is the unique label set, and $W_{j}$ represents weight with every classifier predictions. If weight with every classifier is the same then the above expression can be written as:
 \begin{equation}
     \widehat{y} =\mod(C_1(x),C_2(x),C_3(x), ... C_m(x)) 
 \end{equation}
 In Eq. 2, the $C_i(x)$ represents an individual classifier that predicts labels for a sample $x$. For example, in a binary classification, the mathematical expression representing the model can be written as:
\begin{equation}
  C(x)=\left[  sign\sum_j^mC_{j}(x)  \right]= \left\{
    \begin{array}{ll}
         1 & if \sum_i C_{j}(x) \geq 0\\
        -1 & \mbox{otherwise}\\
    \end{array}
\right.
\end{equation} 
To further elaborate, the above description, lets suppose 5 base classifiers in a binary classification using uniform weights (each base classifier has equal weights).
$C_1(x)\rightarrow 1$, $C_2(x)\rightarrow 1$, $C_3(x)\rightarrow 1$, $C_4(x)\rightarrow 0$ $C_5(x)\rightarrow 0$.
$\hat{y}$= mod $\{$1,1,1,0,0$\}$
$\hat{y}$=1

In the above case, 3 base classifiers out of 5 classify sample $x$ as 1 and 2 classifiers are classified as 0 and then by majority voting the prediction of ensemble model $\hat{y}$ using hard voting will be 1 because maximum classifier is agreed on 1. To further show the superiority of ensemble and voting classifiers, assume $n$ number of based classifiers in the ensemble model. Let us consider that a single classifier has $\varepsilon$ error and further assume that all classifiers are independent. Considering the aforementioned assumptions, the probability error of an ensemble model is represented as follows:
\begin{equation}
 P(y\geq k)=\sum_k^n \binom{n}{k}\varepsilon^{k}(1-\varepsilon)^{n-k}=\varepsilon_{ensemble}   
\end{equation}
\section{Pre-processing of the selected datasets}
This section provides the overview and pre-processing of the selected datasets for the experimental analysis. Based on the cross-comparison of existing datasets performed in Section II, the recently open available Ton-IoT \cite{b23} dataset has been selected for the evaluation of the proposed intrusion detection in IoT networks. This dataset contains 7 different IoT devices namely fridge, motion light, garage door, thermostat, GPS sensor, Modbus, and weather. The data of these datasets are stored separately in 7 CSV file formats. Additionally, the data for the aforementioned heterogeneous IoT devices have been plotted within this section. Figure 3 represents the statistical analysis for the datasets of 2 IoT objects (fridge and garage door) for normal and attack, respectively. Prior to evaluating the proposed model, feature scaling and label encoding has been applied to some IoT data. These datasets have real sensor measurements with low and high values. The statistical analysis for Modbus \& GPS tracker datasets is given in Figure 4 and the dataset statistics of the thermostat, weather sensor, and motion light is shown in Figure 5 with normal and attacked data. The Light Motion sensor dataset has ON/OFF status of lights and similarly, the garage dataset has CLOSE/OPEN in their dataset. We have applied label encoding to prepare it to be used in our proposed model. To evaluate the performance of the Ensemble-based voting classifier on all IoT data, a python script has been implemented to combine all 7 IoT datasets into one CSV file.
Afterward, a median value is chosen for filling the missing variables in the combined IoT dataset as depicted in Figure 2. A recent study reveals that the use of median is less exposed to outlier as compared to the use of mean value, \cite{b40}.  
\begin{table}[!ht]
\centering
\caption{Seven IoT datasets statistics of attack and normal data}
\label{tab:my-table}
\resizebox{\textwidth}{!}{%
\begin{tabular}{|c|c|c|c|c|c|c|c|c|}
\hline
 &  & \multicolumn{7}{c|}{\cellcolor[HTML]{EFEFEF}\textbf{Attacked data}} \\ \cline{3-9} 
\multirow{-2}{*}{\textbf{\begin{tabular}[c]{@{}c@{}}IoT\\ dataset\end{tabular}}} & \multirow{-2}{*}{\textbf{\begin{tabular}[c]{@{}c@{}}Normal\\ data\end{tabular}}} & \textbf{Password} & \textbf{scanning} & \textbf{XSS} & \textbf{DDos} & \textbf{Ransomeware} & \textbf{Injection} & \textbf{Backdoor} \\ \hline
\begin{tabular}[c]{@{}c@{}}Garage\\ Door\end{tabular} & 35000 & 5000 & 529 & 2042 & 5000 & 2902 & 5000 & 5000 \\ \hline
\begin{tabular}[c]{@{}c@{}}Fridge\\ Sensor\end{tabular} & 35000 & 5000 & n/a & 2042 & 5000 & 2902 & 5000 & 5000 \\ \hline
\begin{tabular}[c]{@{}c@{}}GPS\\ tracker\end{tabular} & 35000 & 5000 & 550 & 577 & 5000 & 2833 & 5000 & 5000 \\ \hline
Modbus & 35000 & 5000 & 529 & 577 & n/a & n/a & 5000 & 5000 \\ \hline
\begin{tabular}[c]{@{}c@{}}Light\\ Motion\end{tabular} & 35000 & 5000 & 1775 & 449 & n/a & 2264 & 5000 & 5000 \\ \hline
Weather & 35000 & 5000 & 529 & 866 & 5000 & 2865 & 5000 & 5000 \\ \hline
Thermostat & 35000 & 5000 & 1775 & 449 & 5000 & 2264 & 5000 & 5000 \\ \hline
\end{tabular}%
}
\end{table}

\begin{figure}[] 
\centering
\includegraphics[width=3.5in]{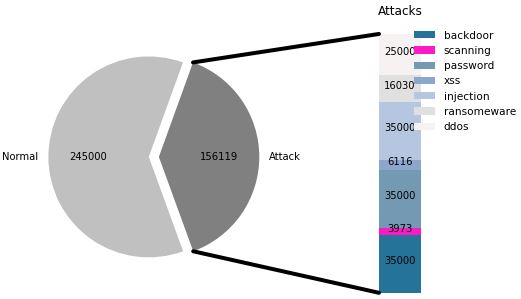}
\caption{Combined IoT dataset statistics}
\label{Courant_2}
\end{figure}
\section{Results and Discussion}
 This section provides a cross-comparison of the reviewed and integrated datasets in our proposed models with different combinations of DT-RF-kNN-NB, DT-RF-NB, and DT-RF-kNN machine learning algorithms. Moreover, the conducted analysis considers four evaluation metrics i.e., accuracy, precision, recall, and F-measure. First, the binary classification for each dataset is given for the combination of algorithms against evaluation matrices. Later, the binary classification is done for combined IoT datasets. 

\begin{table}[]
\centering
\caption{Comparison of the traditional ML algorithms and the proposed models. }
\label{tab:my-table}
\resizebox{\textwidth}{!}{%
\begin{tabular}{|l|c|l|l|l|l|l|l|c|c|c|c|c|}
\hline
\multicolumn{2}{|l|}{} &
  \multicolumn{8}{c|}{{\color[HTML]{000000} \textbf{\begin{tabular}[c]{@{}c@{}}Performance of the state of art machine\\  learning\end{tabular}}}} &
  \multicolumn{3}{c|}{{\color[HTML]{000000} \textbf{\begin{tabular}[c]{@{}c@{}}Our proposed models\\ with different combinations\end{tabular}}}} \\ \hline
\multicolumn{1}{|c|}{{\color[HTML]{000000} \textbf{Datasets}}} &
  \multicolumn{1}{l|}{{\color[HTML]{000000} \textbf{\begin{tabular}[c]{@{}l@{}}Evaluation \\ metrics\end{tabular}}}} &
  {\color[HTML]{000000} \textbf{LR}} &
  {\color[HTML]{000000} \textbf{LDA}} &
  {\color[HTML]{000000} \textbf{kNN}} &
  {\color[HTML]{000000} \textbf{RF}} &
  {\color[HTML]{000000} \textbf{CART}} &
  {\color[HTML]{000000} \textbf{NB}} &
  \multicolumn{1}{l|}{{\color[HTML]{000000} \textbf{SVM}}} &
  \multicolumn{1}{l|}{{\color[HTML]{000000} \textbf{LSTM}}} &
  \textbf{\begin{tabular}[c]{@{}c@{}}(DT-RF-\\ kNN-NB)\end{tabular}} &
  \textbf{\begin{tabular}[c]{@{}c@{}}(DT-RF-\\ NB)\end{tabular}} &
  \multicolumn{1}{l|}{\textbf{\begin{tabular}[c]{@{}l@{}}(DT-RF-\\ kNN)\end{tabular}}} \\ \hline
{\color[HTML]{000000} } &
  {\color[HTML]{000000} \textbf{Accuracy}} &
  0.57 &
  0.77 &
  0.99 &
  0.97 &
  0.97 &
  0.50 &
  0.81 &
  1.00 &
  {\color[HTML]{656565} 1.00} &
  {\color[HTML]{656565} 1.00} &
  {\color[HTML]{656565} 1.00} \\ \cline{2-13} 
{\color[HTML]{000000} } &
  {\color[HTML]{000000} \textbf{Precision}} &
  0.34 &
  0.79 &
  0.99 &
  0.97 &
  0.97 &
  0.53 &
  0.86 &
  1.00 &
  {\color[HTML]{656565} 1.00} &
  {\color[HTML]{656565} 1.00} &
  {\color[HTML]{656565} 1.00} \\ \cline{2-13} 
{\color[HTML]{000000} } &
  {\color[HTML]{000000} \textbf{Recall}} &
  0.58 &
  0.77 &
  0.99 &
  0.97 &
  0.97 &
  0.51 &
  0.82 &
  1.00 &
  {\color[HTML]{656565} 1.00} &
  {\color[HTML]{656565} 1.00} &
  {\color[HTML]{656565} 1.00} \\ \cline{2-13} 
\multirow{-4}{*}{{\color[HTML]{000000} \textbf{Fridge sensor}}} &
  {\color[HTML]{000000} \textbf{F-measure}} &
  0.43 &
  0.77 &
  0.99 &
  0.97 &
  0.97 &
  0.51 &
  0.80 &
  1.00 &
  {\color[HTML]{656565} 1.00} &
  {\color[HTML]{656565} 1.00} &
  {\color[HTML]{656565} 1.00} \\ \hline
\multicolumn{2}{|l|}{{\color[HTML]{000000} }} &
  \multicolumn{8}{l|}{\textbf{}} &
  \multicolumn{3}{l|}{{\color[HTML]{656565} \textbf{}}} \\ \hline
{\color[HTML]{000000} } &
  {\color[HTML]{000000} \textbf{Accuracy}} &
  1.00 &
  1.00 &
  1.00 &
  1.00 &
  1.00 &
  1.00 &
  1.00 &
  1.00 &
  {\color[HTML]{656565} 1.00} &
  {\color[HTML]{656565} 1.00} &
  {\color[HTML]{656565} 1.00} \\ \cline{2-13} 
{\color[HTML]{000000} } &
  {\color[HTML]{000000} \textbf{Precision}} &
  1.00 &
  1.00 &
  1.00 &
  1.00 &
  1.00 &
  1.00 &
  1.00 &
  1.00 &
  {\color[HTML]{656565} 1.00} &
  {\color[HTML]{656565} 1.00} &
  {\color[HTML]{656565} 1.00} \\ \cline{2-13} 
{\color[HTML]{000000} } &
  {\color[HTML]{000000} \textbf{Recall}} &
  1.00 &
  1.00 &
  1.00 &
  1.00 &
  1.00 &
  1.00 &
  1.00 &
  1.00 &
  {\color[HTML]{656565} 1.00} &
  {\color[HTML]{656565} 1.00} &
  {\color[HTML]{656565} 1.00} \\ \cline{2-13} 
\multirow{-4}{*}{{\color[HTML]{000000} \textbf{Garage door}}} &
  {\color[HTML]{000000} \textbf{F-measure}} &
  1.00 &
  1.00 &
  1.00 &
  1.00 &
  1.00 &
  1.00 &
  1.00 &
  1.00 &
  {\color[HTML]{656565} 1.00} &
  {\color[HTML]{656565} 1.00} &
  {\color[HTML]{656565} 1.00} \\ \hline
\multicolumn{2}{|l|}{{\color[HTML]{000000} }} &
  \multicolumn{8}{l|}{} &
  \multicolumn{3}{l|}{{\color[HTML]{656565} }} \\ \hline
{\color[HTML]{000000} } &
  {\color[HTML]{000000} \textbf{Accuracy}} &
  0.86 &
  0.86 &
  0.88 &
  0.85 &
  0.84 &
  0.84 &
  0.86 &
  0.88 &
  {\color[HTML]{656565} 0.96} &
  {\color[HTML]{656565} 0.95} &
  {\color[HTML]{656565} 0.96} \\ \cline{2-13} 
{\color[HTML]{000000} } &
  {\color[HTML]{000000} \textbf{Precision}} &
  0.88 &
  0.88 &
  0.89 &
  0.85 &
  0.85 &
  0.86 &
  0.88 &
  0.89 &
  {\color[HTML]{656565} 0.96} &
  {\color[HTML]{656565} 0.96} &
  {\color[HTML]{656565} 0.96} \\ \cline{2-13} 
{\color[HTML]{000000} } &
  {\color[HTML]{000000} \textbf{Recall}} &
  0.86 &
  0.86 &
  0.88 &
  0.85 &
  0.85 &
  0.85 &
  0.87 &
  0.88 &
  {\color[HTML]{656565} 0.96} &
  {\color[HTML]{656565} 0.96} &
  {\color[HTML]{656565} 0.96} \\ \cline{2-13} 
\multirow{-4}{*}{{\color[HTML]{000000} \textbf{GPS sensor}}} &
  {\color[HTML]{000000} \textbf{F-measure}} &
  0.87 &
  0.87 &
  0.88 &
  0.85 &
  0.85 &
  0.86 &
  0.87 &
  0.88 &
  {\color[HTML]{656565} 0.96} &
  {\color[HTML]{656565} 0.96} &
  {\color[HTML]{656565} 0.96} \\ \hline
\multicolumn{2}{|l|}{{\color[HTML]{000000} }} &
  \multicolumn{8}{l|}{} &
  \multicolumn{3}{c|}{{\color[HTML]{656565} }} \\ \hline
\multicolumn{1}{|c|}{{\color[HTML]{000000} }} &
  {\color[HTML]{000000} \textbf{Accuracy}} &
  0.67 &
  0.67 &
  0.77 &
  0.97 &
  0.98 &
  0.67 &
  0.67 &
  0.67 &
  {\color[HTML]{656565} 0.97} &
  {\color[HTML]{656565} 0.96} &
  {\color[HTML]{656565} 0.96} \\ \cline{2-13} 
\multicolumn{1}{|c|}{{\color[HTML]{000000} }} &
  {\color[HTML]{000000} \textbf{Precision}} &
  0.46 &
  0.46 &
  0.77 &
  0.98 &
  0.99 &
  0.46 &
  0.46 &
  0.47 &
  {\color[HTML]{656565} 0.97} &
  {\color[HTML]{656565} 0.97} &
  {\color[HTML]{656565} 0.97} \\ \cline{2-13} 
\multicolumn{1}{|c|}{{\color[HTML]{000000} }} &
  {\color[HTML]{000000} \textbf{Recall}} &
  0.68 &
  0.68 &
  0.78 &
  0.98 &
  0.98 &
  0.68 &
  0.68 &
  0.68 &
  {\color[HTML]{656565} 0.97} &
  {\color[HTML]{656565} 0.97} &
  {\color[HTML]{656565} 0.97} \\ \cline{2-13} 
\multicolumn{1}{|c|}{\multirow{-4}{*}{{\color[HTML]{000000} \textbf{Modbus}}}} &
  {\color[HTML]{000000} \textbf{F-measure}} &
  0.55 &
  0.55 &
  0.77 &
  0.98 &
  0.99 &
  0.55 &
  0.55 &
  0.55 &
  {\color[HTML]{656565} 0.97} &
  {\color[HTML]{656565} 0.97} &
  {\color[HTML]{656565} 0.97} \\ \hline
\multicolumn{2}{|l|}{{\color[HTML]{000000} }} &
  \multicolumn{8}{l|}{} &
  \multicolumn{3}{l|}{{\color[HTML]{656565} }} \\ \hline
\multicolumn{1}{|c|}{{\color[HTML]{000000} }} &
  {\color[HTML]{000000} \textbf{Accuracy}} &
  0.58 &
  0.58 &
  0.54 &
  0.58 &
  0.58 &
  0.58 &
  \multicolumn{1}{l|}{0.58} &
  \multicolumn{1}{l|}{0.59} &
  {\color[HTML]{656565} 0.58} &
  {\color[HTML]{656565} 0.58} &
  {\color[HTML]{656565} 0.58} \\ \cline{2-13} 
\multicolumn{1}{|c|}{{\color[HTML]{000000} }} &
  {\color[HTML]{000000} \textbf{Precision}} &
  0.34 &
  0.34 &
  0.34 &
  0.34 &
  0.34 &
  0.34 &
  \multicolumn{1}{l|}{0.34} &
  \multicolumn{1}{l|}{0.35} &
  {\color[HTML]{656565} 0.34} &
  {\color[HTML]{656565} 0.35} &
  {\color[HTML]{656565} 0.34} \\ \cline{2-13} 
\multicolumn{1}{|c|}{{\color[HTML]{000000} }} &
  {\color[HTML]{000000} \textbf{Recall}} &
  0.59 &
  0.59 &
  0.59 &
  0.59 &
  0.59 &
  0.59 &
  \multicolumn{1}{l|}{0.59} &
  \multicolumn{1}{l|}{0.59} &
  {\color[HTML]{656565} 0.59} &
  {\color[HTML]{656565} 0.59} &
  {\color[HTML]{656565} 0.58} \\ \cline{2-13} 
\multicolumn{1}{|c|}{\multirow{-4}{*}{{\color[HTML]{000000} \textbf{Light motion}}}} &
  {\color[HTML]{000000} \textbf{F-measure}} &
  0.43 &
  0.43 &
  0.43 &
  0.43 &
  0.43 &
  0.43 &
  \multicolumn{1}{l|}{0.43} &
  \multicolumn{1}{l|}{0.44} &
  {\color[HTML]{656565} 0.43} &
  {\color[HTML]{656565} 0.44} &
  {\color[HTML]{656565} 0.43} \\ \hline
\multicolumn{2}{|l|}{{\color[HTML]{000000} }} &
  \multicolumn{8}{l|}{} &
  \multicolumn{3}{c|}{{\color[HTML]{656565} }} \\ \hline
\multicolumn{1}{|c|}{{\color[HTML]{000000} }} &
  {\color[HTML]{000000} \textbf{Accuracy}} &
  0.66 &
  0.66 &
  0.60 &
  0.66 &
  0.59 &
  0.66 &
  0.66 &
  0.66 &
  {\color[HTML]{656565} 0.66} &
  {\color[HTML]{656565} 0.66} &
  {\color[HTML]{656565} 0.66} \\ \cline{2-13} 
\multicolumn{1}{|c|}{{\color[HTML]{000000} }} &
  {\color[HTML]{000000} \textbf{Precision}} &
  0.44 &
  0.44 &
  0.56 &
  0.59 &
  0.56 &
  0.44 &
  0.44 &
  0.45 &
  {\color[HTML]{656565} 0.78} &
  {\color[HTML]{656565} 0.54} &
  {\color[HTML]{656565} 0.61} \\ \cline{2-13} 
\multicolumn{1}{|c|}{{\color[HTML]{000000} }} &
  {\color[HTML]{000000} \textbf{Recall}} &
  0.66 &
  0.66 &
  0.61 &
  0.66 &
  0.59 &
  0.66 &
  0.66 &
  0.67 &
  {\color[HTML]{656565} 0.66} &
  {\color[HTML]{656565} 0.66} &
  {\color[HTML]{656565} 0.67} \\ \cline{2-13} 
\multicolumn{1}{|c|}{\multirow{-4}{*}{{\color[HTML]{000000} \textbf{Thermostat}}}} &
  {\color[HTML]{000000} \textbf{F-measure}} &
  0.53 &
  0.53 &
  0.57 &
  0.53 &
  0.57 &
  0.53 &
  0.53 &
  0.54 &
  {\color[HTML]{656565} 0.53} &
  {\color[HTML]{656565} 0.53} &
  {\color[HTML]{656565} 0.65} \\ \hline
\multicolumn{2}{|l|}{{\color[HTML]{000000} }} &
  \multicolumn{8}{l|}{} &
  \multicolumn{3}{c|}{{\color[HTML]{656565} }} \\ \hline
\multicolumn{1}{|c|}{{\color[HTML]{000000} }} &
  {\color[HTML]{000000} \textbf{Accuracy}} &
  0.58 &
  0.60 &
  0.81 &
  0.84 &
  0.87 &
  0.59 &
  0.63 &
  0.82 &
  {\color[HTML]{656565} 0.97} &
  {\color[HTML]{656565} 0.97} &
  {\color[HTML]{656565} 0.98} \\ \cline{2-13} 
\multicolumn{1}{|c|}{{\color[HTML]{000000} }} &
  {\color[HTML]{000000} \textbf{Precision}} &
  0.60 &
  0.59 &
  0.81 &
  0.84 &
  0.88 &
  0.72 &
  0.68 &
  0.82 &
  {\color[HTML]{656565} 0.98} &
  {\color[HTML]{656565} 0.97} &
  {\color[HTML]{656565} 0.98} \\ \cline{2-13} 
\multicolumn{1}{|c|}{{\color[HTML]{000000} }} &
  {\color[HTML]{000000} \textbf{Recall}} &
  0.59 &
  0.60 &
  0.81 &
  0.84 &
  0.87 &
  0.59 &
  0.63 &
  0.81 &
  {\color[HTML]{656565} 0.97} &
  {\color[HTML]{656565} 0.97} &
  {\color[HTML]{656565} 0.98} \\ \cline{2-13} 
\multicolumn{1}{|c|}{\multirow{-4}{*}{{\color[HTML]{000000} \textbf{Weather}}}} &
  {\color[HTML]{000000} \textbf{F-measure}} &
  0.53 &
  0.53 &
  0.81 &
  0.84 &
  0.87 &
  0.67 &
  0.55 &
  0.80 &
  {\color[HTML]{656565} 0.97} &
  {\color[HTML]{656565} 0.97} &
  {\color[HTML]{656565} 0.98} \\ \hline
\end{tabular}%
}
\end{table}
The results for binary attack classification are discussed here in this section. The dataset is classified using several traditional algorithms including DT, RF, NB, and the proposed ensemble-based method. The obtained results from the ML algorithms are depicted in Table 3. It is evident from Table 3 that the proposed ensemble-based methods outperform other ML algorithms in most cases. However, in the case of the Modbus and light motion sensor, the accuracy and other accuracy parameters are comparably the same as the ensemble method. Moreover, it is clear from Table 3 that a single combination or a base classifier does not produce accurate results for all sensors.
Data of each sensor are combined into a single CSV file and several experiments were performed. The results of a combined dataset are shown in Table 4. One can see from Table 4 that ensemble-based classifier have optimum accuracy, precision, recall, and F-measure in most cases. However, for the CART-based classifier, we have almost similar results when compared to the ensemble-based classifier. We have also performed experiments for multi-classification and results are depicted in Table 5. It is clear from Table 5 that CART outperforms all other classifiers. 
\begin{table}[]
\centering
\caption{Binary classification on combined IoT dataset}
\label{tab:my-table}
\resizebox{\textwidth}{!}{%
\begin{tabular}{|c|l|c|c|c|c|c|c|c|c|c|c|c|}
\hline
\multicolumn{2}{|l|}{}                                                                                                                                                                    & \multicolumn{8}{c|}{{\color[HTML]{000000} \textbf{Performance of the state of art machine learning}}}                                                                                                                                                                                                                                                                                                                                                                                & \multicolumn{3}{c|}{{\color[HTML]{000000} \textbf{Our proposed models}}}                                                                                                                                                   \\ \hline
{\color[HTML]{000000} \textbf{Datasets}}                                                  & {\color[HTML]{000000} \textbf{\begin{tabular}[c]{@{}l@{}}Evaluation \\ metrics\end{tabular}}} & \multicolumn{1}{l|}{{\color[HTML]{000000} \textbf{LR}}} & \multicolumn{1}{l|}{{\color[HTML]{000000} \textbf{LDA}}} & \multicolumn{1}{l|}{{\color[HTML]{000000} \textbf{kNN}}} & \multicolumn{1}{l|}{{\color[HTML]{000000} \textbf{RF}}} & \multicolumn{1}{l|}{{\color[HTML]{000000} \textbf{CART}}} & \multicolumn{1}{l|}{{\color[HTML]{000000} \textbf{NB}}} & \multicolumn{1}{l|}{{\color[HTML]{000000} \textbf{SVM}}} & \multicolumn{1}{l|}{{\color[HTML]{000000} \textbf{LSTM}}} & \textbf{\begin{tabular}[c]{@{}c@{}}(DT-RF-\\ kNN-NB)\end{tabular}} & \textbf{\begin{tabular}[c]{@{}c@{}}(DT-RF-\\ NB)\end{tabular}} & \multicolumn{1}{l|}{\textbf{\begin{tabular}[c]{@{}l@{}}(DT-RF-\\ kNN)\end{tabular}}} \\ \hline
                                                                                          & \textbf{Accuracy}                                                                             & 0.61                                                    & 0.68                                                     & 0.84                                                     & 0.85                                                    & 0.88                                                      & 0.62                                                    & 0.61                                                     & 0.81                                                      & 0.87                                                               & 0.88                                                           & 0.88                                                                                 \\ \cline{2-13} 
                                                                                          & \textbf{Precision}                                                                            & 0.37                                                    & 0.74                                                     & 0.85                                                     & 0.87                                                    & 0.90                                                      & 0.63                                                    & 0.37                                                     & 0.83                                                      & 0.90                                                               & 0.90                                                           & 0.89                                                                                 \\ \cline{2-13} 
                                                                                          & \textbf{Recall}                                                                               & 0.61                                                    & 0.68                                                     & 0.84                                                     & 0.85                                                    & 0.88                                                      & 0.62                                                    & 0.61                                                     & 0.81                                                      & 0.88                                                               & 0.88                                                           & 0.88                                                                                 \\ \cline{2-13} 
\multirow{-4}{*}{\textbf{\begin{tabular}[c]{@{}c@{}}Combined IoT\\ Dataset\end{tabular}}} & \textbf{F-measure}                                                                            & 0.46                                                    & 0.62                                                     & 0.84                                                     & 0.85                                                    & 0.88                                                      & 0.51                                                    & 0.46                                                     & 0.80                                                      & 0.87                                                               & 0.88                                                           & 0.88                                                                                 \\ \hline
\end{tabular}%
}
\end{table}




\begin{table}[]
\centering
\caption{Multi class classification results on combined IoT dataset}
\label{tab:my-table}
\resizebox{\textwidth}{!}{%
\begin{tabular}{|c|l|c|c|c|c|c|c|c|c|c|c|c|}
\hline
\multicolumn{2}{|l|}{}                                                                                                                                                                    & \multicolumn{8}{c|}{{\color[HTML]{000000} \textbf{Performance of the state of art machine learning}}}                                                                                                                                                                                                                                                                                                                                                                                & \multicolumn{3}{c|}{{\color[HTML]{000000} \textbf{Our proposed models}}}                                                                                                                                                   \\ \hline
{\color[HTML]{000000} \textbf{Datasets}}                                                  & {\color[HTML]{000000} \textbf{\begin{tabular}[c]{@{}l@{}}Evaluation \\ metrics\end{tabular}}} & \multicolumn{1}{l|}{{\color[HTML]{000000} \textbf{LR}}} & \multicolumn{1}{l|}{{\color[HTML]{000000} \textbf{LDA}}} & \multicolumn{1}{l|}{{\color[HTML]{000000} \textbf{kNN}}} & \multicolumn{1}{l|}{{\color[HTML]{000000} \textbf{RF}}} & \multicolumn{1}{l|}{{\color[HTML]{000000} \textbf{CART}}} & \multicolumn{1}{l|}{{\color[HTML]{000000} \textbf{NB}}} & \multicolumn{1}{l|}{{\color[HTML]{000000} \textbf{SVM}}} & \multicolumn{1}{l|}{{\color[HTML]{000000} \textbf{LSTM}}} & \textbf{\begin{tabular}[c]{@{}c@{}}(DT-RF-\\ kNN-NB)\end{tabular}} & \textbf{\begin{tabular}[c]{@{}c@{}}(DT-RF-\\ NB)\end{tabular}} & \multicolumn{1}{l|}{\textbf{\begin{tabular}[c]{@{}l@{}}(DT-RF-\\ kNN)\end{tabular}}} \\ \hline
                                                                                          & \textbf{Accuracy}                                                                             & 0.61                                                    & 0.62                                                     & 0.72                                                     & 0.71                                                    & 0.77                                                      & 0.54                                                    & 0.60                                                     & 0.68                                                      & 0.76                                                               & 0.75                                                           & 0.75                                                                                 \\ \cline{2-13} 
                                                                                          & \textbf{Precision}                                                                            & 0.37                                                    & 0.74                                                     & 0.85                                                     & 0.87                                                    & 0.90                                                      & 0.63                                                    & 0.37                                                     & 0.83                                                      & 0.75                                                               & 0.74                                                           & 0.74                                                                                 \\ \cline{2-13} 
                                                                                          & \textbf{Recall}                                                                               & 0.61                                                    & 0.68                                                     & 0.84                                                     & 0.85                                                    & 0.88                                                      & 0.62                                                    & 0.61                                                     & 0.76                                                      & 0.75                                                               & 0.75                                                           & 0.75                                                                                 \\ \cline{2-13} 
\multirow{-4}{*}{\textbf{\begin{tabular}[c]{@{}c@{}}Combined IoT\\ Dataset\end{tabular}}} & \textbf{F-measure}                                                                            & 0.46                                                    & 0.62                                                     & 0.84                                                     & 0.85                                                    & 0.88                                                      & 0.51                                                    & 0.46                                                     & 0.80                                                      & 0.73                                                               & 0.73                                                           & 0.74                                                                                 \\ \hline
\end{tabular}%
}
\end{table}
\section{Conclusion}
This paper presented a cross-comparison of the available datasets for intrusion attack detection in the Internet of Things (IoT) domain and propose a novel approach based on an ensemble-based voting classifier that combines multiple traditional classifiers as a base learner and gives the vote to the predictions of the traditional classifier in order to get the final prediction. The performance of the proposed model has been compared with the state-of-the-art intrusion detection algorithms available in the current literature and a  comparison has been drawn against the matrices of accuracy, precision, recall, and f-measure. The outcome of the evaluation demonstrated that the proposed strategy outperforms in most of the cases. 
\bibliographystyle{IEEEtran}
\bibliography{ref}
\end{document}